\documentclass[a4paper]{jpconf}
\usepackage{graphicx}
\begin{document}
\title{High-energy astroparticle physics with CALET} 

\author{Paolo Maestro$^1$$^*$}

\address{$^1$ Dept. of Physics, University of Siena, Via Roma 56, 53100 Siena, Italy}
\address{$^*$ on behalf of the CALET collaboration}

\ead{paolo.maestro@pi.infn.it}
\begin{abstract}
The CALorimetric Electron Telescope (CALET) will be installed on the Exposure Facility of the Japanese Experiment Module (JEM-EF) 
on the International Space Station (ISS) in 2014 where it will measure the cosmic-ray fluxes for five years. 
Its main scientific goals are to search for dark matter, 
investigate the mechanism of cosmic-ray acceleration and propagation in the 
Galaxy and discover possible astrophysical sources of high-energy electrons nearby the Earth.
The instrument, under construction, consists of two layers of segmented plastic scintillators 
for the cosmic-ray charge identification (CHD), 
a 3 X$_0$-thick tungsten-scintillating fiber imaging calorimeter (IMC) and a 27 X$_0$-thick lead-tungstate calorimeter (TASC). 
The CHD can provide single-element separation in the interval of atomic number Z from 1 to 40, 
while IMC and TASC can measure the energy of cosmic-ray particles with excellent resolution 
in the range from few GeV up to several hundreds of TeV. Moreover, IMC and TASC 
provide the longitudinal and lateral development of the shower, a key issue for good electron/hadron discrimination. 
In this paper, we will review the status of the mission, the instrument configuration and its expected performance, 
and the CALET capability to measure the different components of the cosmic radiation.
\vspace{-0.8cm}                          
\end{abstract}
\section{Introduction}
CALET (CALorimetric Electron Telescope) is a space-based detector  
developed by a Japanese led international collaboration 
to directly measure the high-energy cosmic radiation on the 
International Space Station (ISS).
CALET is scheduled to  be launched in 2014 by the Japanese rocket 
HTV (H-IIA Transfer Vehicle) and 
robotically installed on the Japanese Experiment Module Exposure Facility (JEM-EF) on ISS.\\
The CALET mission will 
address many of the outstanding questions of High-Energy Astrophysics,
such as the origin of cosmic rays, the mechanism of CR acceleration and galactic propagation,
the existence of dark matter and nearby CR sources, by the observations of 
CR electrons, $\gamma$ rays and nuclei in a wide energy window 
from few GeV up to the TeV region \cite{CALET-GOALS, CALET-DM}.
\section{The CALET instrument and its performance}
\begin{figure}[!h]
\begin{center}
\includegraphics[width=.7\textwidth]{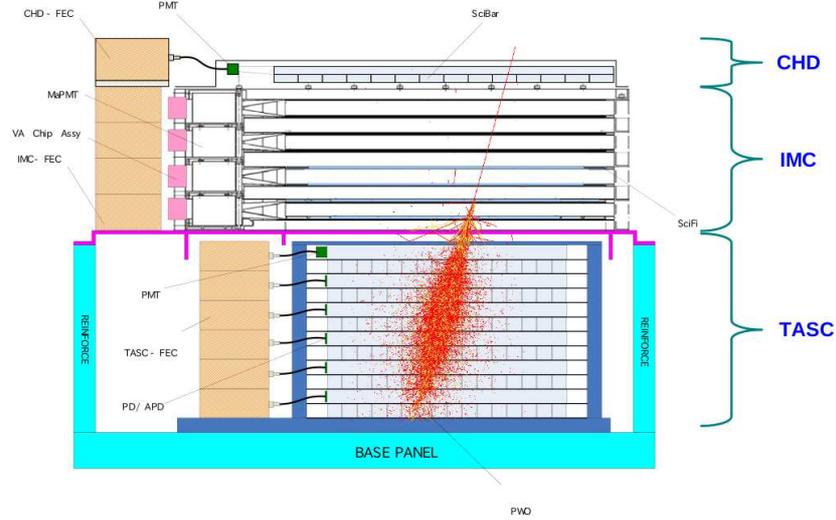}
\caption{Schematic view of CALET. The picture of a simulated shower is superimposed.}
\label{fig1}  
\vspace{-0.7cm}                          
\end{center}
\end{figure}

The CALET instrument consists of a Total AbSorption Calorimeter (TASC), 
a finely segmented  pre-shower IMaging Calorimeter (IMC), 
and a CHarge Detector (CHD) (Fig.~\ref{fig1}).\\
The TASC is a homogeneus calorimeter made of 192 Lead
Tungstate (PWO) ``logs'' (20$\times$19$\times$320 mm$^3$)
arranged in 12 layers. The logs in the top layer are
readout by photomultiplier tubes (PMTs), while  a dual photodiode/avalanche-photodiode
system  is used for the readout of the remaining
layers. 
The TASC can determine the energy of the incident particle 
with excellent energy resolution: $\sim$2\% 
for $e^{\pm}$ and $\gamma$ rays above 100 GeV, $\sim$40\%
for 1 TeV protons and  $\sim$30\% for nuclei at 50 GeV/amu, as estimated from simulations.
Moreover, exploting its shower imaging capabilities, a  proton rejection
$>$10$^5$ can be achieved, 
sufficient to keep
the proton contamination below a few percent
in the observation of CR electrons in the TeV region \cite{CALETMC}.\\
The IMC consists of 7 tungsten plates interleaved with 
double layers of 1 mm$^2$ 
scintillating fibers (SciFi), 
arranged in belts along orthogonal
direction and readout by multianode PMTs, and is capped by
an  additional SciFi layer pair. 
Its surface area is  45$\times$45 cm$^2$ and  its total thickness $\sim$3 radiation lengths ($X_0$). 
The IMC fine granularity  allows to measure precisely the
incident particle trajectory (with angular resolution better than 1$^{\circ}$), 
determine the starting point of the shower and 
separate the incident  from backscattered particles. \\
The charge of the CR nuclei is measured via  the Z$^2$
dependence of the specific ionization loss in a double layered,
segmented, plastic scintillator array (CHD) positioned above the IMC.
Each layer is composed of 14 scintillator paddles (3.2$\times$1.0$\times$44.8 cm$^{3}$) 
each readout by a PMT.
Taking advantage of its excellent charge resolution ($\sim$0.1 electron charge units ($e$) for B, $\sim$0.2$e$ for Fe) \cite{CHD},
CHD can resolve individual chemical elements from Z=1 to Z=40.\\
The total thickness of the instrument is equivalent to
30 $X_0$ and 1.3 proton interaction
length. 
The effective geometrical factor of CALET for high-energy
electrons and nuclei is $\sim$1200 cm$^2$\ sr.
\section{CALET science goals}
It is generally accepted that CRs are
accelerated in shock waves of supernova remnants (SNRs), 
which are the only galactic candidates known with sufficient energy output 
to sustain the CR flux. 
Recent observations of electron synchrotron and gamma-ray emission from SNRs
proved that high-energy charged particles 
are accelerated in SNR shocks  up to energies 
beyond 100 TeV \cite{SNR}.
Unlike the hadronic component of CRs, the electrons, during their diffusion 
in the Galaxy, suffer radiative energy losses
proportional to their squared energy. Thus TeV electrons observed 
at Earth likely originated in sources younger than 10$^5$ years  
and $<$1 kpc far from the Solar System. 
Since the number of such nearby SNRs is limited (e.g.: Vela, Monogem,
Cygnus Loop remnants, and few others), the electron energy spectrum
around 1 TeV could exhibit spectral features
and, at very high energies, a significant
anisotropy in the electron arrival directions 
would be expected. 
Thanks to its excellent energy resolution and capability to discriminate electrons from hadrons,
CALET will be able to investigate possible spectral structures 
by detecting very high-energy electrons 
and possibly provide the 
first experimental evidence of the presence of a nearby CR source.

\begin{figure}[!h]
\begin{center}
\vspace{-0.3cm}                                    
\includegraphics[width=.60\textwidth]{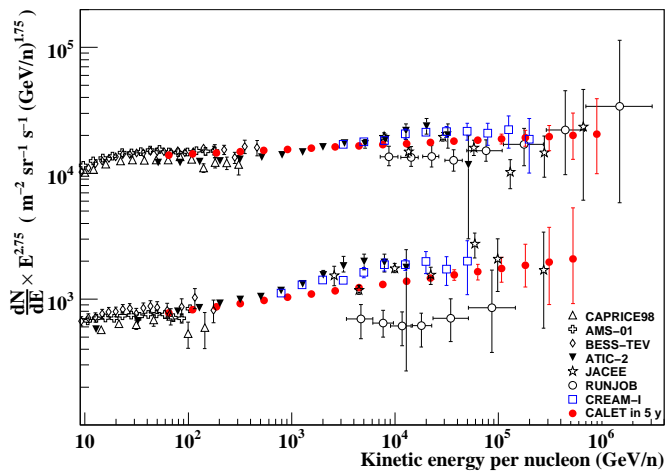}
\caption{
Expected CALET measurement of the energy spectra of proton and He
after 5 years of observation, compared with previous data \cite{AMS, CAPRICE, CREAMp, BESS, ATICp, JACEE, RUNJOB}.
}
\label{fig2}                  
\vspace{-0.4cm}                                    
\end{center}
\end{figure}

\begin{figure}[!h]
\begin{center}
\includegraphics[width=0.9\textwidth]{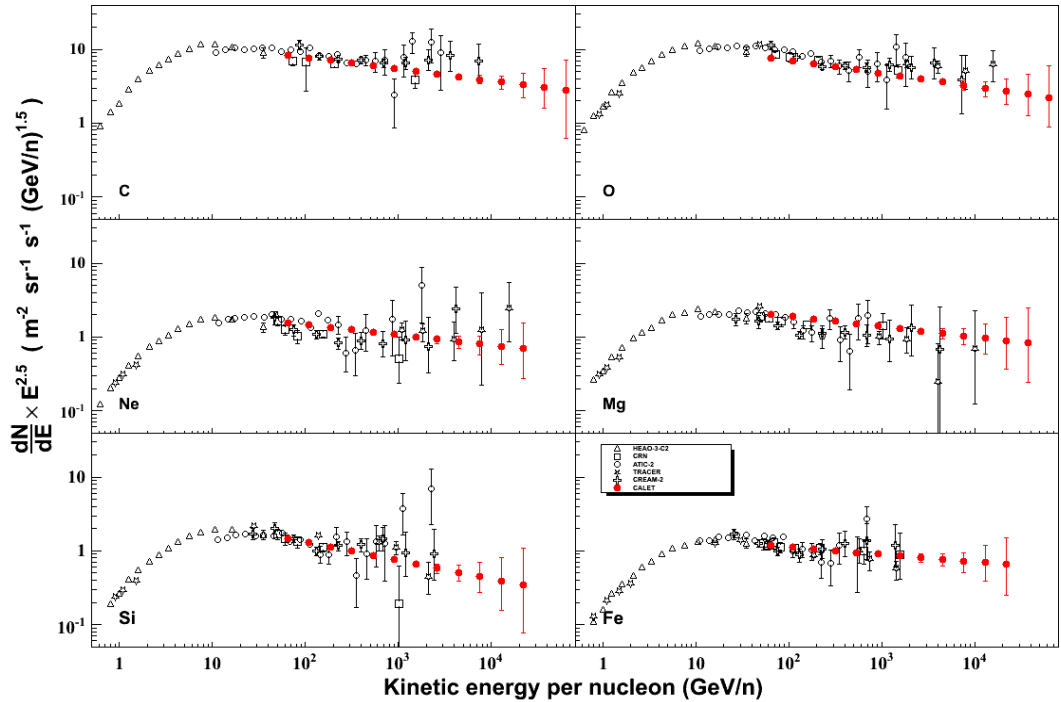}
\caption{Expected CALET measurement of the energy spectra of the more abundant heavy nuclei
after 5 years of observation, compared with previous data \cite{HEAO, CRN, TRACER, ATIC, CREAM}.}
\label{fig3}                            
\vspace{-0.5cm}                                    
\end{center}
\end{figure}

Additional information on the CR acceleration mechanism
might be obtained by directly measuring, besides electrons, the 
 energy spectra of individual CR nuclei up to the PeV scale. 
Possible charge-dependent high-energy spectral
cutoffs, hypothesized to explain the CR ``knee'' \cite{LC},
or spectral hardening 
due to non-linear acceleration mechanisms \cite{Blasi},
could only be investigated by a space experiment
with long enough exposure to extend 
the direct measurement of CR nuclei spectra to unprecedented energies.
CALET  will be able to identify CR nuclei with individual
element resolution and measure their energies in the range
from a few tens of GeV to several hundreds of TeV. 
In five years of data taking on the  ISS, it is expected to 
extend the proton energy spectrum up to $\sim$900 TeV, the He spectrum
up to 400 TeV/amu (Fig.~\ref{fig2}) and 
 measure the energy spectra of the more abundant heavy nuclei C, O,
Ne, Mg, Si and Fe,  with sufficient statistical precision
up to $\sim$20 TeV/amu  (Fig.~\ref{fig3}). It will also investigate precisely possible
spectral features,
like a hardening above 200 GeV/amu recently reported by CREAM \cite{CREAMd},
 or  deviations from a pure power-law spectrum.
Moreover, exploiting the CHD particle identification capability, CALET should 
measure the ultra-heavy ions at few GeV/amu in the 26$<$Z$\le$40 charge range
 with an expected statistics $\sim$5 times larger 
the one collected by the balloon experiment TIGER \cite{CALET-HEAVY}.

The relative  abundances of CR secondary-to-primary
elements (like B/C or sub-Fe/Fe) are known to decrease, following a power-law
in energy $E^{-\delta}$, where $\delta$ is a key parameter in the description of the
CR diffusion in the Galaxy at high energies \cite{CREAMbc}.
At several TeV/amu,
the available data suffer from statistical limitations and 
 large systematic errors, due
to the residual atmospheric overburden at balloon altitude,
and has not allowed so far
to place a stringent experimental constraint on the value of  $\delta$.
Taking advantage of its long exposure in space and the absence
of atmosphere, CALET will provide new data to improve
the accuracy of the present measurements  above
100 GeV/amu and extend them above 1 TeV/amu.

Besides studying the CR sources and diffusion, 
CALET will also conduct a sensitive search for signatures
of dark matter candidates (e.g.: Weakly Interacting Massive
Particles (WIMPs), Kaluza-Klein particles, etc.) in both the electron and gamma-ray spectra. 
With its excellent energy resolution and long exposure in space, 
it will be able to detect possible 
lines due to WIMP decays in the gamma-ray spectrum above few hundreds of GeV, 
and shed light on the controversial anomalous excess in the electron spectrum recently
reported by the balloon experiments ATIC \cite{ATICe} but not confirmed by FERMI \cite{FERMI}.

Finally, additional CALET science objectives
are the detailed study of the solar modulation by the measurement of 
the electron spectrum  time evolution below 10 GeV, 
and the detection of gamma-ray bursts  and X-ray transients 
 by means of a dedicated scintillator-based Gamma-ray Burst Monitor 
associated to the main CALET telescope.


\vspace{-0.3cm}                          
\section*{References}
\begin{thebibliography}{9}
\bibitem{CALET-GOALS} Yoshida K {\it et al.} 2011 {\it Proc. of  31$^{st}$ ICRC (Lodz)} vol 6 p 360 
\bibitem{CALET-DM} Torii S {\it et al.} 2011 {\it Nucl. Instr. And Meth.} A {\bf 630} 55
\bibitem{CALETMC} Akaike Y  {\it et al.} 2010 {\it Adv. Space Res.} {\bf 45} 690  
\bibitem{CHD} Marrocchesi P S {\it et al.} 2011 {\it Nucl. Instr. And Meth.} A {\bf 659} 477         
\bibitem{SNR} Aharonian F A {\it et al.} 2004 {\it Nature} {\bf 432}  75
\bibitem{LC} Lagage P O and Cesarsky C J 1983 {\it A\&A} {\bf 125} 249 
\bibitem{Blasi} P.~Blasi {\it et al.} 2012 {\it Phys.Rev.Lett.}  {\bf 109}  061101 
\bibitem{CREAMd} Ahn H S {\it et al.} 2010 {\it ApJ} {\bf 714}  L89            
\bibitem{CALET-HEAVY} Rauch B {\it et al.} 2011 {\it Proc. of 31$^{st}$ ICRC (Lodz)} vol 6 p 348 
\bibitem{AMS} Aguilar M {\it et al.} 2002 {\it Phys. Rep.} {\bf 366}  331
\bibitem{CAPRICE} Boezio M {\it et al.} 1999 {\it ApJ} {\bf 518}  457
\bibitem{CREAMp} Ahn H S {\it et al.} 2011 {\it ApJ}  {\bf 728} (2011) 122   
\bibitem{BESS} Haino S {\it et al.} 2004 {\it Phys. Lett.} B {\bf 594} 35                        
\bibitem{ATICp} Panov A D {\it et al.} 2009 {\it Bull. Russ. Acad. Sci. Phys.} {\bf 73} 564
\bibitem{JACEE} Asakimori K {\it et al.} 1998 {\it ApJ} {\bf 502}  278
\bibitem{RUNJOB} Derbina V A {\it et al.} 2005 {\it ApJ} {\bf 628} L41
\bibitem{HEAO} Engelmann J J {\it et al.} 1990 {\it A\&A} {\bf 233}  96
\bibitem{CRN} M\"{u}ller D {\it et al.} 1991 {\it ApJ} {\bf 374}   356
\bibitem{TRACER} Ave M {\it et al.} 2008 ApJ {\bf 678}  262 
\bibitem{ATIC} Panov A D {\it et al.} 2006 {\it Adv. Space Res.} {\bf 37} 1944
\bibitem{CREAM} Ahn H S {\it et al.} 2009 {\it ApJ}  {\bf 707}   593-603                           
\bibitem{CREAMbc} Ahn H S {\it et al.} 2008 {\it Astropart. Phys.}  {\bf 30}  133-141                    
\bibitem{ATICe} Chang J {\it et al.} 2008 {\it Nature} {\bf 456}  362
\bibitem{FERMI} Abdo A A {\it et al.} 2009 {\it Phys. Rev. Lett.} {\bf 102}  181101
\end{thebibliography}
\end{document}